\documentclass[aps,prb,reprint,groupedaddress,showpacs]{revtex4-1}

\usepackage{graphicx}
\usepackage{natbib}
\usepackage{amsmath}

\begin{document}


\title{Reversible Electric-Field Driven Magnetic Domain Wall Motion}


\author{K\'{e}vin J. A. Franke,$^{1}$ Ben Van de Wiele,$^{2}$ Yasuhiro Shirahata,$^{3}$ Sampo J. H\"{a}m\"{a}l\"{a}inen,$^{1}$ Tomoyasu Taniyama,$^{3}$ and Sebastiaan van Dijken$^{1}$}
\email[]{sebastiaan.van.dijken@aalto.fi}
\affiliation{$^{1}$NanoSpin, Department of Applied Physics, Aalto University School of Science, P.O. Box 15100, FI-00076 Aalto, Finland.}
\affiliation{$^{2}$Department of Electrical Energy, Systems and Automation, Ghent University, Ghent B-9000, Belgium.}
\affiliation{$^{3}$Materials and Structures Laboratory, Tokyo Institute of Technology, 4259 Nagatsuta, Midori-ku, Yokohama, Japan.}


\date{\today}

\begin{abstract}
Control of magnetic domain wall motion by electric fields has recently attracted scientific attention because of its potential for magnetic logic and memory devices. Here, we report on a new driving mechanism that allows for magnetic domain wall motion in an applied electric field without the concurrent use of a magnetic field or spin-polarized electric current. The mechanism is based on elastic coupling between magnetic and ferroelectric domain walls in multiferroic heterostructures. Pure electric-field driven magnetic domain wall motion is demonstrated for epitaxial Fe films on BaTiO$_3$ with in-plane and out-of-plane polarized domains. In this system, magnetic domain wall motion is fully reversible and the velocity of the walls varies exponentially as a function of out-of-plane electric field strength.  
\end{abstract}


\maketitle

\section{introduction}
Domain walls in ferromagnetic thin films or nanowires are conventionally driven by magnetic fields or spin-polarized electric currents \cite{1974JAP....45.5406S,1984JAP....55.1954B,1999Sci...284..468O,2003ApPhL..83..509G,2004PhRvL..92h6601T,2005NatMa...4..741B,2005EL.....69..990T,2005PhRvL..94j6601K,2006PhRvL..96s7207H,2008ApPhL..93z2504M}. The velocity of magnetic domain walls varies with the driving force and various material properties (magnetic anisotropy, damping constant, saturation magnetization etc.), which gives rise to several dynamic regimes \cite{2007AnRMS..37..415K}. In the thermally activated creep regime, domain wall motion depends sensitively on the disorder-induced pinning energy barrier and the depinning field \cite{1998PhRvL..80..849L}. This notion has led to various demonstrations of electric-field control over the pinning strength and velocity of magnetic domain walls via voltage-induced changes of magnetic anisotropy. Examples include the use of dielectric gates or ferroelectric films to manipulate the magnetic anisotropy via charge modulation or band shifting \cite{2012ApPhL.100s2408B,2012NatCo...3E.847S,2012NatCo...3E.888C,2012ApPhL.101q2403B,2012PhRvB..86w5130M,2013ApPhL.102l2406B,2013ApPhL.103j2411F,2013ApPhL.103v2902H}. Other promising methods utilize strain coupling to piezoelectric materials \cite{2013NatCo...4E1378L,2013NatMa..12..808D} or electric-field induced ionic diffusion \cite{2013NatNa...8..411B}. Voltage-controlled domain wall gates and traps based on these concepts provide new prospects for magnetic logic and memory technologies. In all instances, the magnetic domain walls are driven by a magnetic field or electric current and the velocity of the walls is altered by an electric-field effect on the magnetic anisotropy. Besides, electric-field induced magnetic domain wall deformations\cite{2008ApPhL..92k2509C} and magnetic switching via lateral domain wall motion\cite{2012ApPhL.101g2402P} have been demonstrated in piezoelectric-magnetostrictive heterostructures with competing anisotropies. Full electric-field control over the position and velocity of magnetic domain walls, however, has thus far remained elusive.   

\begin{figure*}
\includegraphics{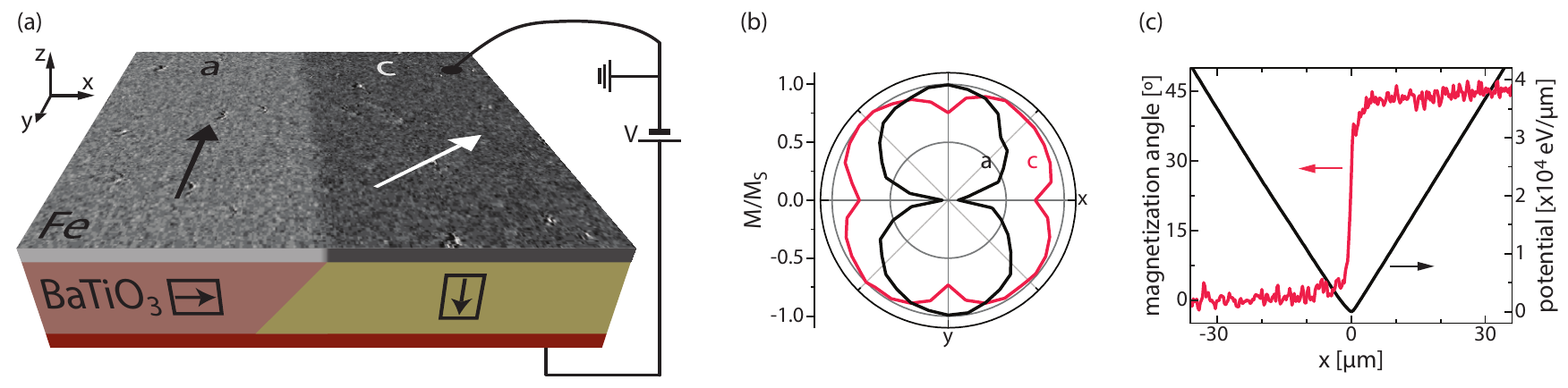}
\caption{\label{fig1} (a) Schematic illustration and Kerr microscopy image of a Fe/BaTiO$_3$ heterostructure with a ferroelectric $a$ and $c$ domain. Arrows indicate the direction of ferroelectric polarization in the BaTiO$_3$ substrate and the direction of magnetization in the Fe film in zero magnetic field. Measurements of electric-field driven magnetic domain wall motion are conducted by contacting the Fe/Au layer to ground and applying bias voltage pulses to the back side of the BaTiO$_3$ substrate. (b) Polar plots of the remanent magnetization on top of the $a$ domain (black line) and $c$ domain (red line). (c) Measured profile of the magnetic domain wall in (a) (red line) and calculated pinning potential (black line).}
\end{figure*}

In multiferroic heterostructures magnetoelectric coupling between a ferroelectric material and a ferromagnetic film has been extensively studied during the last decade and various interaction mechanisms have been identified as promising routes towards exclusively electric-field controlled magnetism \cite{2006Natur.442..759E,2007NatMa...6...21R,ADMA:ADMA201003636,0953-8984-24-33-333201}. In exchange- and strain-coupled systems, direct correlations between the polarization direction within ferroelectric domains and the locally-induced magnetic anisotropy have been demonstrated \cite{2008NatMa...7..478C,2009PhRvL.103y7601L,ADMA:ADMA201100426,2011PhRvL.107u7202H,2012PhRvB..86a4408C,2013PhRvB..87e4410S,2014PhRvL.112a7201F}. Multiferroic heterostructures with lateral anisotropy modulations are characterized by strong pinning of magnetic domain walls onto narrow ferroelectric domain boundaries \cite{2012PhRvB..85i4423F}. This robust coupling effect forces the magnetic domain walls to follow their ferroelectric counterparts when the latter are displaced in an applied electric field, thus providing a new way to drive magnetic domain walls by pure electrical means. In our previous work, we studied electric-field induced magnetic switching via lateral domain wall motion in polycrystalline CoFe films that were strain-coupled to the ferroelastic $a_1$ and $a_2$ domains of a BaTiO$_3$ substrate \cite{2012NatSR...2E.258L}. Due to partial strain transfer during film growth and the in-plane orientation of ferroelectric polarization in these structures, complex domain patterns evolved in an out-of-plane electric field and control over domain wall motion could not be obtained.    

Here, we report on the ability to reversibly and deterministically drive magnetic domain walls by an electric field only. Full electric-field control over the magnetic domain wall velocity in the absence of a magnetic field or spin-polarized electric current is demonstrated using epitaxial Fe films on single-crystal BaTiO$_3$ substrates with alternating in-plane ($a$ domains) and out-of-plane ($c$ domains) ferroelectric polarization. The magnetic domain walls of the Fe film are strongly pinned onto the $a-c$ boundaries of the BaTiO$_3$ substrate by abrupt changes in the symmetry and strength of magnetic anisotropy. Back-and-forth motion of a ferroelectric $a-c$ boundary and a pinned magnetic domain wall is realized by the application of positive and negative out-of-plane voltage pulses. In our proof-of-concept experiments, the domain wall velocity is varied over five orders of magnitude by adjusting the electric field strength. Micromagnetic simulations indicate that near-180$^\circ$ transverse magnetic domain walls can be stabilized in magnetic nanowires on top of BaTiO$_3$. The spin structure of such domain walls is protected against dynamic breakdown and strong elastic pinning to $a-c$ boundaries is sustained up to high driving velocities.  

\section{sample fabrication and characterization}

Fe films with a thickness of 20 nm were grown onto single-crystal BaTiO$_3$ substrates with a thickness of 0.5 mm using molecular beam epitaxy at 300$^\circ$C. Regular ferroelectric $a$ and $c$ domains were imprinted into the Fe layer upon cooling through the Curie temperature of BaTiO$_3$ at $T_C$ = 120$^\circ$C, which coincides with a cubic-to-tetragonal structural phase transition. At room temperature, the Fe film was capped by 5 nm of Au to prevent oxidation during sample characterization. Growth of Fe onto the (001)-oriented BaTiO$_3$ substrates was epitaxial with a Fe[110]//BaTiO$_3$[100] crystal alignment in the film plane \cite{2012ApPhL.101z2405L}.

The ferromagnetic domains and domain walls of the Fe film were imaged using a magneto-optical Kerr effect microscope. Magnetization reversal on top of the ferroelectric $a$ and $c$ domains was characterized for in-plane magnetic fields. Electric-field driven magnetic domain wall motion in zero magnetic field was analyzed in the same microscope by applying out-of-plane voltage pulses across the BaTiO$_3$ substrate. In these experiments, the backside of the sample was contacted by double-sided copper tape and wire bonding to the metallic Fe/Au layer was used to create the top electrode. Bias voltage pulses were generated by a bipolar power supply with the Fe/Au top electrode contacted to ground. The strength of the electric field was varied from 0.5 kV/cm to 8 kV/cm. Magnetic domain wall displacements were determined from Kerr microscopy images that were recorded prior to and immediately after the application of a voltage pulse. 
 
\section{experimental results}

At room temperature, the tetragonal BaTiO$_3$ substrate consists of domains with in-plane ($a$ domain) and out-of-plane ($c$ domain) ferroelectric polarization. Rotation of the polarization at the domain boundaries coincides with an abrupt change of the BaTiO$_3$ in-plane lattice structure: the unit cell of $a$ domains in the (001)-oriented surface is rectangular with a tetragonality of 1.1\% and the in-plane lattice of $c$ domains is cubic. The change in lattice symmetry is reflected by the magnetic anisotropy of the Fe film whose magnetoelastic component is induced via strain transfer at the heterostructure interface and inverse magnetostriction. Because of the negative magnetostriction constant of Fe, the easy axis of magnetization on top of the ferroelectric $a$ domains is oriented prependicular to the ferroelectric polarization and thus parallel to the domain wall, while the two magnetic easy axes on top of the $c$ domains are aligned at 45$^\circ$. Figure \ref{fig1}(a) depicts the experimental configuration. Here, a schematic illustration of an $a$ and $c$ domain in the BaTiO$_3$ substrate is combined with a Kerr microscopy image of the magnetic microstructure in the Fe film for zero magnetic field. Full angular analyses of the local remanent magnetization (polar plots in Fig. \ref{fig1}(b)) confirm the twofold and fourfold symmetry of magnetic anisotropy on top of the ferroelectric $a$ and $c$ domains. The corresponding magnetic anisotropy strengths, as determined from fits to local hard-axis hysteresis curves, are $K_{a}=2\times10^4$ J/m$^3$ and $K_{c}=1\times10^4$ J/m$^3$.

\begin{figure}
\includegraphics{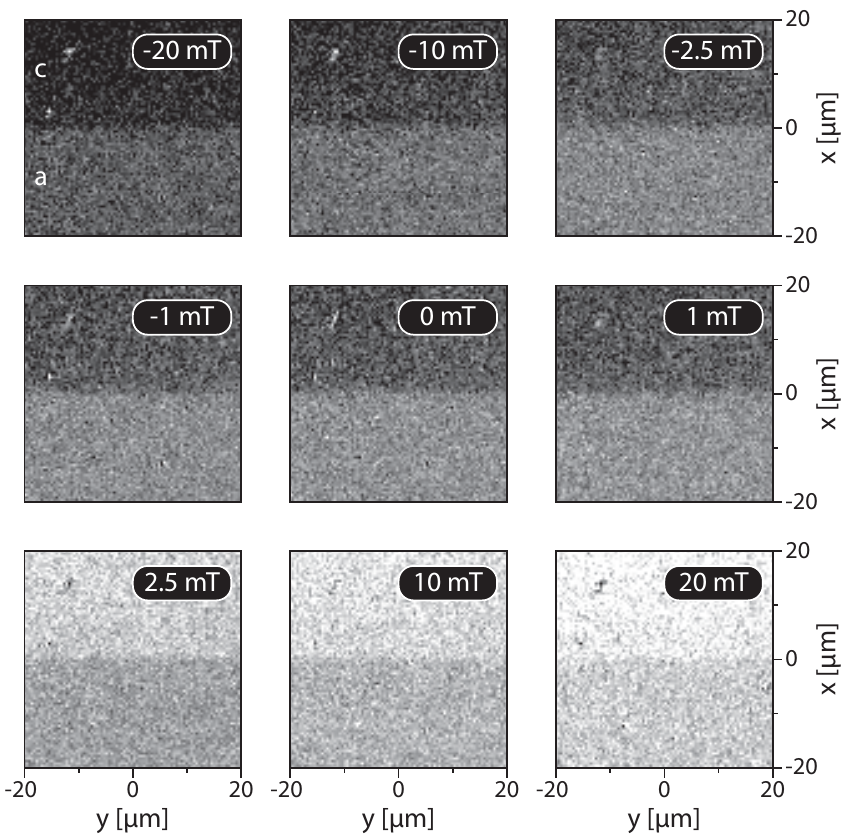}
\caption{\label{fig2}Kerr microscopy images of magnetization reversal in a Fe film on top of a BaTiO$_3$ substrate. The in-plane magnetic field and axis of magneto-optical Kerr effect contrast are aligned perpendicular to the domain wall.} 
\end{figure}

\begin{figure*}
\includegraphics{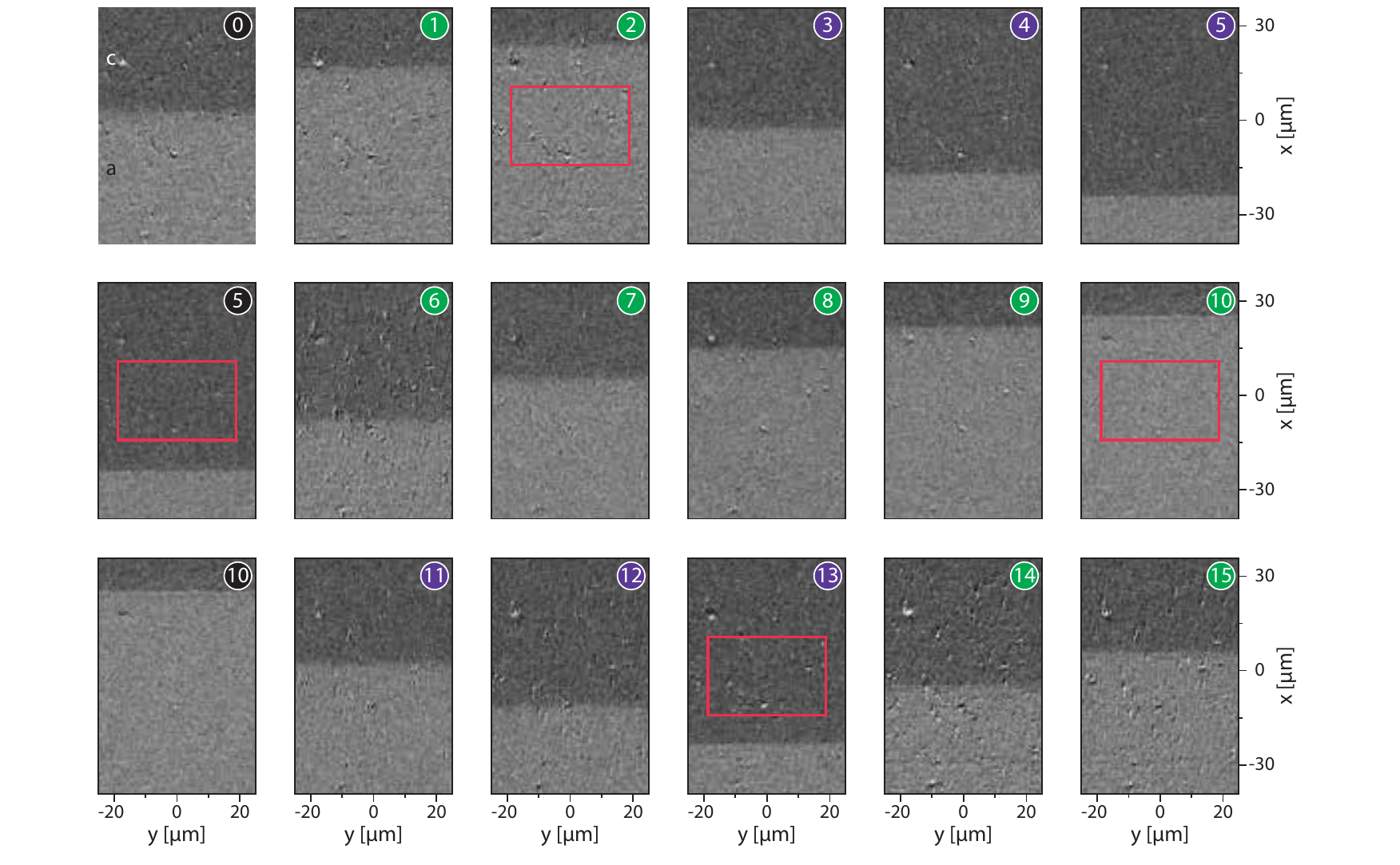}
\caption{\label{fig3}Kerr microscopy images illustrating the motion of a magnetic domain wall in a Fe film during the application of positive (green circles) and negative (violet circles) voltage pulses across the BaTiO$_3$ substrate. Black circles indicate that no voltage pulse was applied between two consecutive images. The pulse strength is $|V|=100$ V, which corresponds to an electric field of $|E|=2$ kV/cm. The red boxes denote an area for which magnetic hysteresis curves were measured after electric field driven magnetic domain wall motion (see Fig. \ref{fig4}(b)).} 
\end{figure*}

\subsection{Magnetic domain wall pinning}

The abrupt change in symmetry and strength of magnetic anisotropy on top of BaTiO$_3$ $a-c$ domain walls produces narrow and straight anisotropy boundaries inside the Fe film. Magnetic domain walls are strongly pinned by these anisotropy boundaries. This pinning effect can be understood by considering a lateral displacement of a pinned magnetic domain wall away from a ferroelectric boundary. Such a movement would cause a misalignment between the local magnetization and the easy anisotropy axes, thereby enhancing the total magnetic anisotropy energy of the system. The pinning potential can be calculated if the magnetization profile of the domain wall and the symmetry and strength of the magnetic anisotropy on either side of the anisotropy boundary are known (see Appendix for details). Figure \ref{fig1}(c) shows the magnetization profile of a pinned magnetic domain wall on top of a ferroelectric $a-c$ boundary as measured by Kerr microscopy. Also shown is the pinning potential, which is calculated using the wall profile and the experimental values for $K_{a}$ and $K_{c}$. The minimum of the pinning potential is located on top of the ferroelectric $a-c$ boundary ($x=0$) and the magnetic anisotropy energy increases sharply when the distance between the two ferroic domain walls is enhanced. We note that this analysis of magnetic domain wall pinning must be considered as a first order approximation, not taking into account possible deformations of the magnetic domain wall profile and assuming a perfectly abrupt ferroelectric boundary. The latter simplification is justified by experimental results and theoretical calculations indicating a ferroelastic domain wall width of only $2-5$ nm in BaTiO$_3$ \cite {1992ApPhL..60..784Z,2006PhRvB..74j4104H,2006ApPhL..89r2903Z}, which is more than one order of magnitude smaller than the width of magnetic domain walls in 20 nm thick Fe films. 

Strong magnetic domain wall pinning is confirmed by the Kerr microscopy data of Fig. \ref{fig2}. In the images, magnetization reversal in the Fe film for a magnetic field perpendicular to the domain wall is shown. The wall that separates the two ferromagnetic domains is fully immobilized by elastic coupling to the underlying ferroelectric $a-c$ boundary in the  BaTiO$_3$ substrate. The magnetization of both domains reverse independently. The magnetic contrast on top of the ferroelectric $a$ domain (lower part of the images) changes gradually, indicating coherent magnetization rotation. This behavior is explained by a perpendicular alignment of the easy anisotropy axis and the in-plane magnetic field. Magnetization reversal on top of the ferroelectric $c$ domain (upper part of the images) proceeds by coherent rotation and abrupt switching at a magnetic field of 2.5 mT. In this case, the magnetic field is oriented along one of the hard axes of the cubic magnetic anisotropy. While the spin rotation of the magnetic domain wall changes with applied magnetic field strength, its position remains firmly fixed on top of the ferroelectric boundary in the BaTiO$_3$ substrate. 

\begin{figure*}
\includegraphics{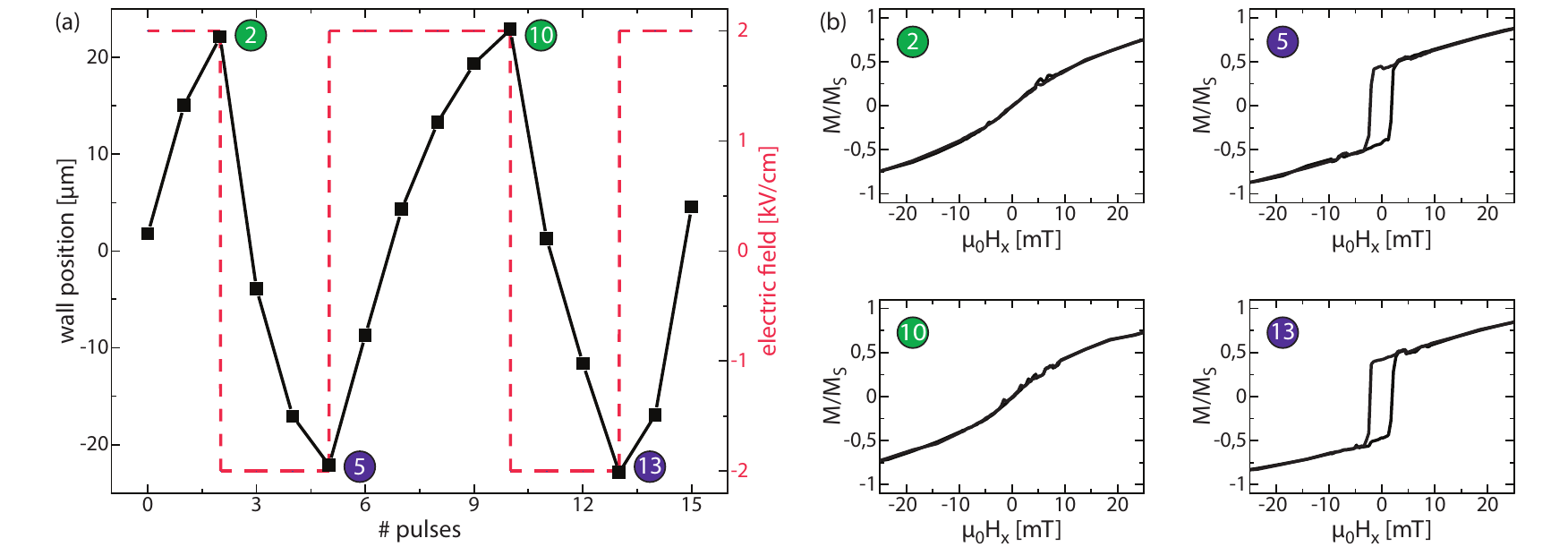}
\caption{\label{fig4}(a) Magnetic domain wall position in the Fe film after the application of a series of positive and negative electric field pulses across the BaTiO$_3$ substrate. The polarity and strength of the electric field is indicated by a dashed red line. The numbers correspond to the measurements in Fig. \ref{fig3}. (b) Magnetic hysteresis curves for a selected sample area (red box in Fig. \ref{fig3}). The data evidence that alternations between uniaxial ($a$ domain) and biaxial ($c$ domain) magnetic anisotropy during back and forth motion of the magnetic domain wall are reproducible.}   
\end{figure*}

\subsection{Electric-field driven magnetic domain wall motion}

Domain boundaries in the BaTiO$_3$ substrate can be moved by the application of an electric field. The concurrent motion of magnetic anisotropy boundaries and their pinning potential in an elastically coupled ferromagnetic film can be utilized to drive magnetic domain walls by pure electrical means. Reversible motion of an electric-field driven magnetic domain wall is demonstrated in Fig. \ref{fig3}. The Kerr microscopy images show the remanent magnetic microstructure of the Fe film on top of a BaTiO$_3$ substrate with a ferroelectric $a-c$ domain boundary. Reversible motion of the ferroelectric boundary and the pinned magnetic domain wall is achieved by the application of out-of-plane electric field pulses (see Fig. \ref{fig1}(a)). If the electric field is aligned along the direction of ferroelectric polarization in the $c$ domain (negative bias voltage), the $c$ domain grows at the expense of the neighboring $a$ domain by lateral wall motion towards the bottom of the images. A positive bias voltage, on the other hand, shrinks the $c$ domain by moving the ferroelectric boundary and pinned magnetic domain wall back up. Strong elastic coupling necessitates that the magnetic domain wall in the Fe film closely follows the displacement of the ferroelectric domain boundary.  

The position of the magnetic domain wall as a function of the number of electric field pulses is plotted in Fig. \ref{fig4}(a). The velocity of the magnetic domain wall decreases when it departs from its original position by several micrometers. This feature is characteristic for ferroelastic domain wall motion in BaTiO$_3$ and other ferroelectric materials \cite{1953PhRv...90..193M,Tagantsev}. The built-in restoring force that the $a-c$ domain wall experiences is caused by a repulsive interaction between ferroelastic walls in the BaTiO$_3$ substrate. Since the magnetic domain wall in the Fe film is strongly pinned onto the ferroelastic boundary, its electric field driven motion is intrinsically linked to the ferroelectric sub-system. Isolation of a single domain wall in multiferroic heterostructures would eliminate the dependence of domain wall motion on sample position.   

\begin{figure}
\includegraphics{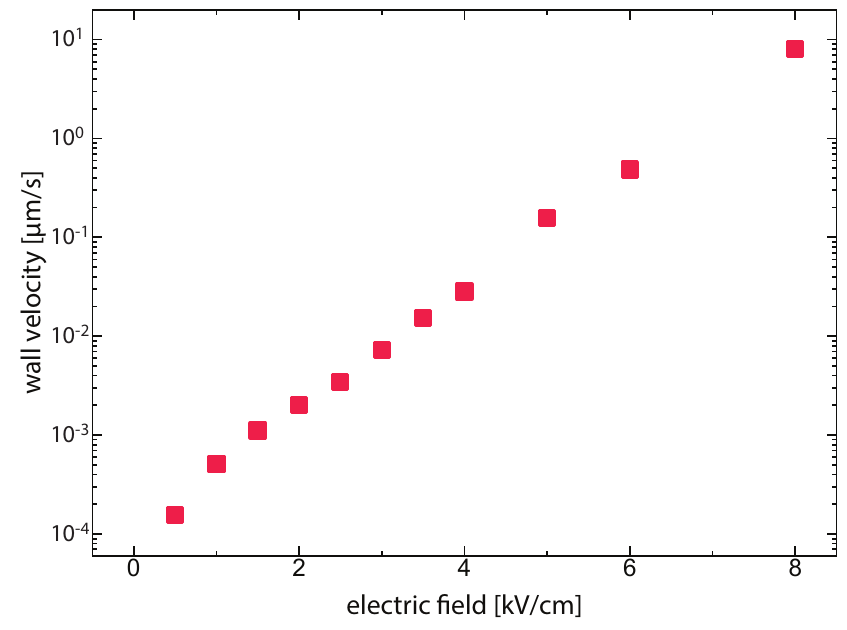}
\caption{\label{fig5} Magnetic domain wall velocity as a function of applied electric field.}   
\end{figure}

Figure \ref{fig4}(b) shows magnetic hysteresis curves from the same sample area (red box in Fig. \ref{fig3}) after repeated back and forth motion of the magnetic domain wall. The direction of the in-plane magnetic field in these measurements is perpendicular to the domain wall. The data indicate an alternation of magnetic anisotropy from uniaxial (when the entire area is located on top of the ferroelectric $a$ domain) to biaxial (when the underlying ferroelectric domain is of $c$ type). The strength and symmetry of the magnetic anisotropy, however, are preserved upon repeated domain wall cycling in an applied electric field. Thus, contrary to other electric field effects \cite{2012ApPhL.100s2408B,2012NatCo...3E.847S,2012NatCo...3E.888C,2012ApPhL.101q2403B,2013ApPhL.102l2406B,2013ApPhL.103j2411F,2013ApPhL.103v2902H,2012PhRvB..86w5130M,2013NatCo...4E1378L,2013NatMa..12..808D,2013NatNa...8..411B,2008ApPhL..92k2509C,2012ApPhL.101g2402P}, the mechanism discussed in this paper does not alter the magnitude of magnetic anisotropy. Instead, the magnetic domain wall in the Fe film follows the sideways motion of an anisotropy boundary that is induced by a ferroelectric $a-c$ domain wall in the BaTiO$_3$ substrate, but the distinctive magnetic anisotropy on top of the $a$ and $c$ domains is conserved.

The dependence of the average domain wall velocity on electric field strength was recorded by adjusting the duration of the electric field pulse ($\Delta$$t$) to the electric field strength ($E$), so that accurate measurements of domain wall displacement ($\Delta$$x$) could be obtained by Kerr microscopy. The velocity of elastically coupled magnetic and ferroelectric domain walls was subsequently inferred from $v$($E$) = $\Delta$$x$/$\Delta$$t$. The results are presented in Fig. \ref{fig5}. An exponential increase of domain wall velocity with increasing electric field is obtained, which is characteristic of thermally assisted depinning of ferroelectric domain walls in the strong pinning regime \cite{1953PhRv...90..193M}. The velocity of the magnetic domain wall in the Fe film can be controlled over five orders of magnitude in the current experiments. While this result is very promising, the maximum recorded velocity falls short of the anticipated upper limit for electric-field driven ferroelectric domain walls. The main reason for this is the thickness of the BaTiO$_3$ substrate (0.5 mm), which requires the use of large bias voltages. An extension of this proof-of-concept investigation to higher domain wall velocities could be attained by replacing the BaTiO$_3$ substrate by a thinner ferroelectric crystal or film. Such configurations would allow for the application of short pulses of large electric field at low bias voltage. Since the dynamics of electric-field driven magnetic domain wall motion is mainly governed by the ferroelectric sub-system, the upper bound of fast domain wall motion is set by the ferroelectric material. Studies on domain wall dynamics in BaTiO$_3$ indicate that wall velocities up to 1000 m/s can be achieved at room temperature \cite{1963JAP....34.3255S}, which is comparable to fast domain wall motion in magnetic nanowires \cite{1999Sci...284..468O,2005NatMa...4..741B,2006PhRvL..96s7207H,2008ApPhL..93z2504M}.

\begin{figure*}
\includegraphics{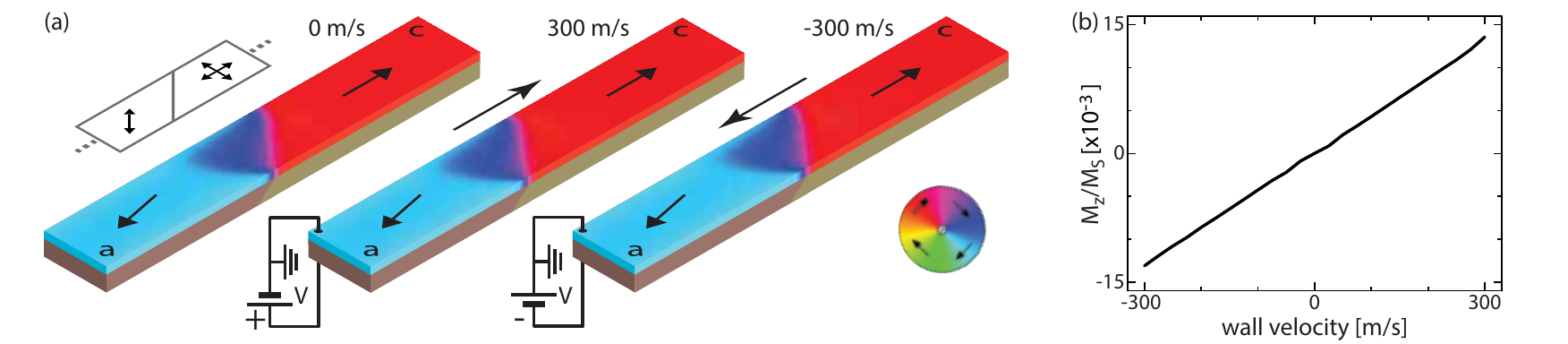}
\caption{\label{fig6} Micromagnetic simulations of magnetic domain wall motion in a 200 nm wide and 5 nm thick Fe wire. The schematic in the upper left corner indicates the easy magnetic anisotropy axes in the nanowire. The shape anisotropy of the wire stabilizes a near-180$^\circ$ transverse magnetic domain wall in zero magnetic field, which is strongly pinned by an abrupt anisotropy boundary on top of a ferroelectric $a-c$ domain wall. Fast motion of the anisotropy boundary in both directions results in an equally fast displacement of the transverse magnetic domain wall with only minor deformations of the internal wall structure. The data in (a) illustrate the magnetic structure of the nanowire at rest and for $v_{AB}=+300$ m/s and $v_{AB}=-300$ m/s. The evolution of a small out-of-plane magnetization component with increasing domain wall velocity is shown in (b).}
\end{figure*}

\section{micromagnetic simulations}

Micromagnetic simulations were performed to study the structure and motion of pinned magnetic domain walls in patterned nanowires. The simulations were conducted with the graphics processing unit (GPU) based micromagnetic simulator MuMax \cite{2011JMMM..323.2585V}. To closely mimic strain coupling in the Fe/BaTiO$_3$ system, experimental values for the magnetic anisotropy on top of ferroelectric $a$ and $c$ domains were used (i.e. $K_{a}=2\times10^4$ J/m$^3$ and $K_{c}=1\times10^4$ J/m$^3$). Other input parameters for the Fe film included a saturation magnetization of $M_{S}=1.7\times10^6$ A/m, an exchange constant of $K_{ex}=2.1\times10^{-11}$ J/m, and a damping constant of $\alpha$ = 0.01. An infinitely long 200 nm wide nanowire with a thickness of 5 nm and successive $a$ and $c$ domains of length 3.2 $\mu$m was considered. Using periodic boundary conditions along the nanowire axis, the computational area was restricted to one set of $a$ and $c$ domains. Finite difference cells of size 3.125 nm $\times$ 3.125 nm $\times$ 5 nm were used to discretize the geometry. Motion of the anisotropy boundary at a velocity $v_{AB}$ was implemented by shifting the anisotropy boundary over a distance of one cell ($\delta$$x$ = 3.125 nm) every $\delta$$t$ = $\delta$$x$/$v_{AB}$.

The simulations reveal that patterning of ferromagnetic films into nanowire geometries offers additional degrees of freedom for the engineering of robust and strongly pinned magnetic domain walls. For example, despite the 45$^\circ$ angle between the magnetic anisotropy axes on top of the ferroelectric $a$ and $c$ domains, near-180$^\circ$ transverse magnetic domain walls can be stabilized in magnetic nanowires because of competing shape anisotropy as shown in Fig. \ref{fig6}. At rest and zero applied magnetic field, a transverse domain wall is formed and pinned onto the magnetic anisotropy boundary between the $a$ and $c$ domains. Fast motion of the anisotropy boundary to the right or left by 300 m/s does not depin the magnetic domain wall. Only a small out-of-plane magnetization component develops as a function of wall velocity (Fig. \ref{fig6}(b)), which is reminiscent of pre-Walker breakdown behavior of magnetic field or current-driven domain wall motion \cite{2005EL.....69..990T}. Strong dynamic deformations of the internal wall structure, however, do not occur during the electric-field driven process. The absence of pronounced dynamic instabilities can be ascribed to the abrupt anisotropy boundary, which provides topological protection against breakdown \cite{2014ApPhL.104a2401V}. 

\section{conclusion}

Reversible electric-field driven magnetic domain wall motion is demonstrated for Fe/BaTiO$_3$ heterostructures. Strong elastic pinning of magnetic domain walls onto ferroelectric domain boundaries forms the basis of an all-electrical driving mechanism, which allows for accurate control over the position and velocity of magnetic domain walls. Reproducible back-and-forth motion of a magnetic domain wall is obtained by the use of a ferroelectric $a-c$ domain boundary and out-of-plane electric field pulses. Future experiments can build on these results by exploring various magnetic domain wall structures and higher electric fields in patterned nanowires.             

\section{appendix}
\begin{figure}
\includegraphics{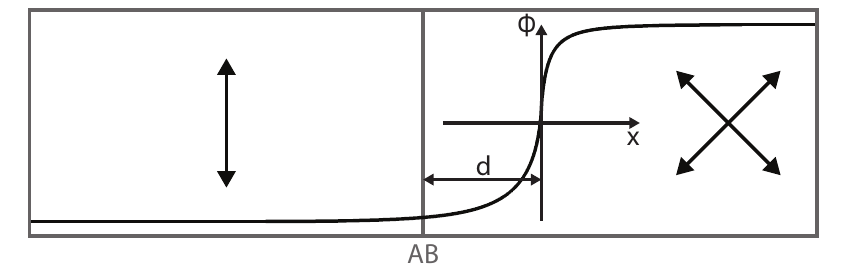}
\caption{\label{fig7} The domain wall pinning potential $V(d)$ of an anisotropy boundary (AB) between domains with uniaxial and cubic magnetic anisotropy can be estimated by calculating the increase of anisotropy energy when the magnetic domain wall ($\phi(x)$) is moved to position $d$. The anisotropy boundary in the Fe film is created by elastic coupling to an underlying ferroelectric $a-c$ domain wall in the BaTiO$_3$ substrate.}
\end{figure}

The domain wall pinning potential that is created in a ferromagnetic film through elastic coupling to a ferroelectric domain wall can be estimated by considering a lateral displacement of the magnetic domain wall profile with respect to the ferroelectric boundary (see Fig. \ref{fig7}). In equilibrium, the magnetization profile of a domain wall $\phi(x)$ that separates two domains $R_i$ ($i$ = 1,2) with different magnetic anisotropy is given by total energy minimization. If this wall profile is assumed to be constant, the pinning potential in zero magnetic field can be estimated by considering magnetic anisotropy only. For domains $R_i$ exhibiting both uniaxial ($K_{ui}$) and cubic ($K_{ci}$) anisotropy, the energy density can be written as:

\begin{equation}
e_{i} = K_{ui}\sin^2\left(\phi(x)-\theta_{i}\right)+K_{ci}\sin^2\left(2\left(\phi(x)-\psi_{i}\right)\right),
\label{eq1}
\end{equation}
\\
where $\theta_{i}$ and $\psi_{i}$ indicate the direction of the easy anisotropy axes with respect to the domain wall. Sideways motion of the magnetic domain wall profile by a distance $d$ away from the ferroelectric domain boundary increases the anisotropy energy of the system. The increase of magnetic energy per unit domain wall length for a ferromagnetic film of thickness $t$ is given by:

\begin{widetext}
\begin{align}
E(d) = t \cdot\int_{-\infty}^{-d}\!dxK_{u1}\sin^2\left(\phi(x)-\theta_{1}\right)
+K_{c1}\sin^2\left(2\left(\phi(x)-\psi_{1}\right)\right) \nonumber \\
+t \cdot\int_{-d}^{\infty}\!dxK_{u2}\sin^2\left(\phi(x)-\theta_{2}\right)
+K_{c2}\sin^2\left(2\left(\phi(x)-\psi_{2}\right)\right) 
\label{eq2}
\end{align}
\\
In our experimental system, domain $R_1$ corresponds to the uniaxial $a$ domain with $\theta_1$ = 0$^\circ$ and domain $R_2$ corresponds to the cubic $c$ domain with $\psi_2$ = 45$^\circ$. For this configuration, Eq.~\ref{eq2} simplifies to:
\\
\begin{align}
E(d) = t \cdot\int_{-\infty}^{-d}\!dxK_{a}\sin^2\left(\phi(x)\right)
+t \cdot\int_{-d}^{\infty}\!dxK_{c}\cos^2\left(2\phi(x)\right) 
\label{eq3}
\end{align}
\\
The domain wall pinning potential at position $d$ is given by $V(d)=E(d)-E(d=0)$, i.e.:
\\
\begin{align}
V(d) = t \cdot\left[\int_{-\infty}^{-d}\!dxK_{a}\sin^2\left(\phi(x)\right)
+\int_{-d}^{\infty}\!dxK_{c}\cos^2\left(2\phi(x)\right)\right] \nonumber \\
-t \cdot\left[\int_{-\infty}^{0}\!dxK_{a}\sin^2\left(\phi(x)\right)
+\int_{0}^{\infty}\!dxK_{c}\cos^2\left(2\phi(x)\right)\right]
\label{eq4}
\end{align}
\\
In our Fe/BaTiO$_3$ samples, $K_a=2K_c$, which results in a nearly symmetric domain wall pinning potential (see Fig. \ref{fig1}(c)). \\ \\
\end{widetext}

%

\begin{acknowledgments}
We are thankful to T.H.E. Lahtinen and A. Casiraghi for fruitful discussions. This work was supported by the European Research Council (ERC-2012-StG 307502-E-CONTROL), the Industrial Technology Research Grant Program in 2009 from NEDO of Japan, JSPS KAKENHI (Grant No. 24.7390), the Advanced Materials Development and Integration of Novel Structured Metallic and Inorganic Materials Project of MEXT, and the Collaborative Research Project of the Materials and Structures Laboratory, Tokyo Institute of Technology. K.J.A.F. acknowledges financial support from the Finnish Doctoral Program in Computational Sciences and B.V.d.W. received financial support from the Flanders Research Foundation. 
\end{acknowledgments}


%

\end{document}